\documentclass[a4paper,12pt]{article}
\usepackage{amsmath}
\usepackage{accents}
\usepackage[english]{babel}
\usepackage{graphicx}
\usepackage[usenames]{color}
\usepackage{colortbl}

\begin{document}

\centerline {\LARGE{Quantum state geometry and entanglement of two spins}}
\centerline {\LARGE{with anisotropic interaction in evolution}}
\medskip
\centerline {A. R. Kuzmak}
\centerline {\small \it E-Mail: andrijkuzmak@gmail.com}
    \medskip
\centerline {\small \it Department for Theoretical Physics, Ivan Franko National University of Lviv,}
\medskip
\centerline {\small \it 12 Drahomanov St., Lviv, UA-79005, Ukraine}

{\small

Quantum evolution of a two-spin system with anisotropic Heisenberg Hamiltonian in the magnetic field is considered.
We show that this evolution happens on some manifold with geometry depending on the ratio between the interaction
couplings and on the initial state. The Fubini-Study metric of this manifold is calculated. The entanglement of the states belonging
to this manifold is examined. Also we investigate similar problem for a two-spin system described by the Dzyaloshinsky-Moria Hamiltonian.
The problem is solved by using the fact that this Hamiltonian and the anisotropic Heisenberg Hamiltonian are linked by the unitary transformation.

\medskip

}

\section{Introduction\label{sec1}}

Understanding of the geometry of quantum state manifolds is useful in the study of many problems related to the quantum evolution.
For example, using the fact that the whole space of states of a two-level system is represented by the Bloch sphere, we easily obtain
the trajectory of quantum evolution between two states of this system (see, for instance, \cite{FSM2,ref1,ref2,ref3,TOSTSS,FSM,TMTSPMF}).
This trajectory is a curve between two points on the sphere. The minimal trajectory of a spin-$1/2$ particle,
driven by a magnetic field, was found, using geometrical properties of the quantum state space \cite{FSM,TMTSPMF}). Also, in a similar way,
the quantum brachistochrone problem for an arbitrary spin in a
magnetic field \cite{brachass} and the Zermelo navigation problem \cite{ZNP1,ZNP2,ZNP3,ZNP4,ZNP5} were solved. Another well-known problem,
where the understanding of the geometry of the quantum states space plays an important role, is the problem of finding the time-optimal
Hamiltonian which provides the evolution between two specified quantum states (the quantum brachistochrone problem) \cite{OHfST}.
This problem was solved by using symmetry properties of the quantum state space. Also it was shown that the problem of
finding of the quantum circuit of unitary operations which provide time-optimal evolution on a system of qubits \cite{OCGQC,GAQCLB,QCAG,QGDM} and qutrits
\cite{GQCQ} is related to the problem of finding of the minimal distance between two points on a Riemannian metric.
One can find more about geometrical properties of quantum systems in the papers \cite{FSM2,FSM,BSSQWF,BVQUDIT,GQM,BVNLS}.

The information about the geometry of the quantum state manifolds can be obtained by examination of the Fubini-Study metric of these manifolds
\cite{FSM2, FSM, FSM3, FSM0, FSM1, FSM4}. For instance, the Fubini-Study metric of the quantum evolution submanifold of the
projective Hilbert space which is associated with the evolution generated by a set of independent operators was studied in \cite{FSM3}.
The metric of the ground state manifold of the $XY$ chain in a transverse magnetic field was obtained in \cite{FSM4}.
Also with help of the Fubini-Study metric the geometrical properties of some well-known coherent state manifolds
\cite{RSMQS, CSRS} and the rotational manifold \cite{brachass} were investigated.

In our previous paper \cite{torus} we considered the quantum evolution of a two-spin system described by the isotropic Heisenberg Hamiltonian
in the magnetic field. The metric of the manifold defined by the states which can be achieved during such evolution was obtained.
In the present paper, we investigate evolution of a two-spin system with anisotropic Heisenberg interaction which is placed in the magnetic
field directed along the $z$-axis (section \ref{sec2}). In the section \ref{sec3} it is shown that this evolution happens on
a manifold with geometry depending on the ratio between the interaction couplings and on the initial state.
The entanglement of the states belonging to this manifold is examined in the section \ref{sec4}. Also a similar problem is studied in the case
of the Dzyaloshinsky-Moria (DM) interaction between spins (section \ref{sec5}). Conclusions are presented in the section \ref{sec6}.

\section{The quantum evolution of two spins with an anisotropic Heisenberg interaction \label{sec2}}

We consider a two-spin system represented by an anisotropic Heisenberg Hamiltonian in the external magnetic
field directed along the $z$-axis. The Hamiltonian of the system is as follows
\begin{eqnarray}
H=H_{xx}+H_{zz}+H_{mf},
\label{form1}
\end{eqnarray}
with
\begin{eqnarray}
&&H_{xx}=J\left(\sigma_x^1\sigma_x^2+\sigma_y^1\sigma_y^2\right),\label{form1_1}\\
&&H_{zz}=\alpha J \sigma_z^1\sigma_z^2,\label{form1_2}\\
&&H_{mf}=h_z\left(\sigma_z^{1}+\sigma_z^{2}\right),\label{form1_4}
\end{eqnarray}
where $\sigma_i^1=\sigma_i\otimes 1$,
$\sigma_i^2=1\otimes\sigma_i$, and $\sigma_i$ are the Pauli matrices,
$J$ is the interaction coupling, $h_z$ is proportional to the
strength of the magnetic field and $\alpha$ is any real number that defines the anisotropy of the system.
When $\alpha=1$ then interaction between two spins is represented by the isotropic Heisenberg Hamiltonian.
In the case of $\alpha=0$ the interaction between two spins is described by the Heisenberg $XX$ model.
The Hamiltonian (\ref{form1}) has four eigenvalues, namely, $\alpha J+2h_z$, $\alpha J-2h_z$, $-\alpha J+2J$, and $-\alpha J-2J$
with the corresponding eigenvectors
\begin{eqnarray}
&&\vert\uparrow\uparrow\rangle,\label{form2_1}\\
&&\vert\downarrow\downarrow\rangle,\label{form2_2}\\
&&\frac{1}{\sqrt{2}}\left(\vert\uparrow\downarrow\rangle + \vert\downarrow\uparrow\rangle\right),\label{form2_3}\\
&&\frac{1}{\sqrt{2}}\left(\vert\uparrow\downarrow\rangle - \vert\downarrow\uparrow\rangle\right).
\label{form2_4}
\end{eqnarray}

Let us consider the quantum evolution of a two-spin system with Hamiltonian (\ref{form1}). Using the fact that
$H_{xx}$, $H_{zz}$ and $H_{mf}$ commute between themselves the evolution operator can be represented by the following expression
\begin{eqnarray}
U(t)=e^{-iH_{xx}t}e^{-iH_{zz}t}e^{-ih_z\sigma_z^1t}e^{-ih_z\sigma_z^2t},
\label{form3}
\end{eqnarray}
where
\begin{eqnarray}
&&e^{-iH_{xx}t}=1+\left[\cos\left(2J t\right)-1\right]\frac{1}{2}\left(1-\sigma_z^1\sigma_z^2\right)-i\frac{\sin\left(2J t\right)}{2J}H_{xx},\label{form3_1}\\
&&e^{-iH_{zz}t}=\cos\left(\alpha Jt\right)-i\frac{\sin\left(\alpha Jt\right)}{\alpha J}H_{zz}.\label{form3_2}
\end{eqnarray}
Here we use the fact that
\begin{eqnarray}
H_{xx}^{2n}=\left(2J\right)^{2n}\frac{1}{2}\left(1-\sigma_z^1\sigma_z^2\right),\quad H_{xx}^{2n+1}=\left(2J\right)^{2n}H_{xx},\quad H_{zz}^2=\left(\alpha J\right)^2,\nonumber
\end{eqnarray}
where $n=1,2,3,\ldots$ We set $\hbar=1$, which means that the
energy is measured in the frequency units. In the basis labelled by
$\vert\uparrow\uparrow\rangle$, $\vert\uparrow\downarrow\rangle$,
$\vert\downarrow\uparrow\rangle$ and
$\vert\downarrow\downarrow\rangle$, the evolution operator
$U(t)$ can be represented as
\begin{eqnarray}
U(t)=\left( \begin{array}{ccccc}
e^{-i\left(2h_z+\alpha J\right)t} & 0 & 0 & 0\\
0 & \cos\left(2J t\right)e^{i\alpha Jt} & -i\sin\left(2J t\right)e^{i\alpha Jt} & 0 \\
0 & -i\sin\left(2J t\right)e^{i\alpha Jt} & \cos\left(2J t\right)e^{i\alpha Jt} & 0 \\
0 & 0 & 0 & e^{i\left(2h_z-\alpha J\right)t}
\end{array}\right).
\label{form4}
\end{eqnarray}

An arbitrary quantum state of two spins can be expressed in the following form
\begin{eqnarray}
\vert\psi_i\rangle=a\vert\uparrow\uparrow\rangle+b\vert\uparrow\downarrow\rangle+c\vert\downarrow\uparrow\rangle+d\vert\downarrow\downarrow\rangle,
\label{form4_1}
\end{eqnarray}
where $a$, $b$, $c$ and $d$ are the complex parameters which satisfy the normalization condition
$\vert a\vert^2+\vert b\vert^2+\vert c\vert^2+\vert d\vert^2=1$.
Let us consider the evolution of two spins having started from state (\ref{form4_1})
with parameters $a=a_i$, $b=b_i$, $c=c_i$, and $d=d_i$. The action of the evolution operator (\ref{form3})
on this state is as follows
\begin{eqnarray}
&&\vert\psi(\theta,\phi)\rangle=e^{i\frac{\alpha\theta}{2}}\left[a_ie^{-i\left(\phi+\alpha\theta\right)}\vert\uparrow\uparrow\rangle + \left(b_i\cos\theta-ic_i\sin\theta\right)\vert\uparrow\downarrow\rangle \right.\nonumber\\
&&\left. + \left(-ib_i\sin\theta+c_i\cos\theta\right)\vert\downarrow\uparrow\rangle + d_ie^{i\left(\phi-\alpha\theta\right)}\vert\downarrow\downarrow\rangle\right],
\label{form5}
\end{eqnarray}
where
\begin{eqnarray}
\theta=2Jt,\quad \phi=2h_zt.
\label{form5_1}
\end{eqnarray}

It is worth mentioning that the interaction coupling $J$ is fixed, so parameters $t$ and $h_z$ are free.
Thus the state (\ref{form5}) is defined by two real parameters $\theta$ and $\phi$ which in turn are defined by the value of the magnetic field $h_z$
and the period of evolution $t$. For any pre-defined set of parameters $\theta$ and $\phi$ there exists a set of values $h_z$ and $t$.

The state (\ref{form5}) satisfies some periodic conditions on the parameters $\theta$ and $\phi$. These conditions depend on the initial
parameters $a_i$, $b_i$, $c_i$, $d_i$ and parameter $\alpha$. Let us consider possible cases of these conditions in detail. So, there are
the following cases:
\begin{enumerate}
\item If $a_i=d_i=0$ and $b_i\neq \pm c_i$, the state (\ref{form5}) takes the form
\begin{eqnarray}
\vert\psi(\theta)\rangle=e^{i\frac{\alpha\theta}{2}}\left[\left(b_i\cos\theta-ic_i\sin\theta\right)\vert\uparrow\downarrow\rangle + \left(-ib_i\sin\theta+c_i\cos\theta\right)\vert\downarrow\uparrow\rangle\right].
\label{form5_2}
\end{eqnarray}
As we can see, this state depends only on parameter $\theta$.
It is easy to see that it satisfies the following periodic condition
\begin{eqnarray}
\vert\psi(\theta+\pi)\rangle=-e^{i\frac{\alpha\pi}{2}}\vert\psi(\theta)\rangle.
\label{form5_3}
\end{eqnarray}
\item If $a_i\neq d_i\neq0$ and $b_i= c_i=0$, we obtain the following state
\begin{eqnarray}
\vert\psi(\phi)\rangle=e^{-i\frac{\alpha\theta}{2}}\left[a_ie^{-i\phi}\vert\uparrow\uparrow\rangle + d_ie^{i\phi}\vert\downarrow\downarrow\rangle\right],
\label{form5_4}
\end{eqnarray}
which depends on parameter $\phi$ and satisfies the following condition
\begin{eqnarray}
\vert\psi(\phi+\pi)\rangle=-\vert\psi(\phi)\rangle.
\label{form5_5}
\end{eqnarray}
We can see that parameter $\theta$ affects only on the phase of the state (\ref{form5_4}).
\item If $\alpha=1$, $c_i=b_i$ or $\alpha=-1$, $c_i=-b_i$ and at least one of the parameters
$a_i$ or $d_i$ is non-zero, we have
\begin{eqnarray}
\vert\psi(\phi)\rangle=e^{\mp i\frac{\theta}{2}}\left[a_ie^{-i\phi}\vert\uparrow\uparrow\rangle + b_i\left(\vert\uparrow\downarrow\rangle  \pm \vert\downarrow\uparrow\rangle\right) + d_ie^{i\phi}\vert\downarrow\downarrow\rangle\right],
\label{form5_6}
\end{eqnarray}
where upper sign corresponds to the case of $\alpha=1$ and lower sign corresponds to the case of $\alpha=-1$. The state (\ref{form5_6})
also depends on the parameter $\phi$ and parameter $\theta$ affects only on phase of state. This state satisfies the following periodic condition
\begin{eqnarray}
\vert\psi(\phi+2\pi)\rangle=\vert\psi(\phi)\rangle.
\label{form5_7}
\end{eqnarray}
\item If $\alpha\neq 1$, $c_i=b_i$ or $\alpha\neq -1$, $c_i=-b_i$ and also at least one of the parameters $a_i$ or $d_i$ is non-zero,
we have
\begin{eqnarray}
\vert\psi(\theta,\phi)\rangle=e^{i\frac{\alpha\theta}{2}}\left[a_ie^{-i\left(\phi+\alpha\theta\right)}\vert\uparrow\uparrow\rangle + b_ie^{\mp i\theta}\left(\vert\uparrow\downarrow\rangle  \pm \vert\downarrow\uparrow\rangle\right) + d_ie^{i\left(\phi-\alpha\theta\right)}\vert\downarrow\downarrow\rangle\right].
\label{form5_8}
\end{eqnarray}
Here upper sign corresponds to the case of $\alpha\neq 1$ and lower sign corresponds to the case of $\alpha\neq-1$.
So, we obtain the following periodic conditions
\begin{eqnarray}
&&\vert\psi(\theta,\phi+2\pi)\rangle=\vert\psi(\theta,\phi)\rangle,\nonumber\\
&&\vert\psi(\theta+\frac{\pi}{\alpha\mp1},\phi+\pi)\rangle=-e^{-i\frac{\alpha\pi}{2(\alpha\mp 1)}}\vert\psi(\theta,\phi)\rangle.
\label{form5_9}
\end{eqnarray}
\item If $c_i\neq\pm b_i$ and $\alpha=p/q$ is a rational number, where $p$ and $q$ are coprime integers,
then it is easy to see from (\ref{form5}) that the following equalities are satisfied
\begin{eqnarray}
\vert\psi(\theta,\phi+2\pi)\rangle=\vert\psi(\theta,\phi)\rangle
\label{form5_10}
\end{eqnarray}
and
\begin{eqnarray}
\vert\psi(\theta+q\pi,\phi)\rangle=e^{-i\frac{p\pi}{2}}\vert\psi(\theta,\phi)\rangle,
\label{form5_11}
\end{eqnarray}
for the case of $p$ and $q$ are both odd or
\begin{eqnarray}
\vert\psi(\theta+q\pi,\phi+\pi)\rangle=e^{-i\left(\frac{p}{2}+1\right)\pi}\vert\psi(\theta,\phi)\rangle.
\label{form5_12}
\end{eqnarray}
for the case of $p$ is even and $q$ is odd and vice versa.
\item In the case if $c_i\neq\pm b_i$ and $\alpha$ is an irrational number we have only one periodic condition for the state (\ref{form5})
\begin{eqnarray}
\vert\psi(\theta,\phi+2\pi)\rangle=\vert\psi(\theta,\phi)\rangle.
\label{form5_13}
\end{eqnarray}
\end{enumerate}

So, analyzing these particular cases we can conclude that in the first three cases the evolution of the system happens on one-parametric closed manifolds.
In cases 4 and 5 the manifolds are two-parametric and closed. In the last case the manifold is two-parametric and opened through parameter $\theta$.
Let us find the geometry of these manifolds.

\section{The Fubini-Study metric of a two-spin quantum state manifold \label{sec3}}

The Fubini-Study metric is defined by the infinitesimal distance $ds$ between two neighbouring pure quantum states $\vert\psi(\xi^{\mu})\rangle$ and
$\vert\psi(\xi^{\mu}+d\xi^{\mu})\rangle$ \cite{FSM2, FSM, FSM0, FSM1}
\begin{eqnarray}
ds^2=g_{\mu\nu}d\xi^{\mu} d\xi^{\nu},
\label{form10}
\end{eqnarray}
where $\xi^{\mu}$ is a set of real parameters which define the state $\vert\psi(\xi^{\mu})\rangle$. The components of the metric tensor
$g_{\mu\nu}$ have the form
\begin{eqnarray}
g_{\mu\nu}=\gamma^2\Re\left(\langle\psi_{\mu}\vert\psi_{\nu}\rangle-\langle\psi_{\mu}\vert\psi\rangle\langle\psi\vert\psi_{\nu}\rangle\right),
\label{form11}
\end{eqnarray}
where $\gamma$ is an arbitrary factor which is often chosen to have value of $1$, $\sqrt{2}$ or $2$ and
\begin{eqnarray}
\vert\psi_{\mu}\rangle=\frac{\partial}{\partial\xi^{\mu}}\vert\psi\rangle.
\label{form12}
\end{eqnarray}

Let us calculate the metrics of the manifolds defined by the states (\ref{form5}). Using the fact that this state is determined by
the two parameters $\theta$ and $\phi$  we obtain the following scalar products
\begin{eqnarray}
&&\langle\psi\vert \psi_{\theta}\rangle=-i\left[\left(\alpha-1\right)A+1-B\right],\quad \langle\psi\vert \psi_{\phi}\rangle=-iD,\nonumber\\
&&\langle\psi_{\theta} \vert \psi_{\theta}\rangle=\left(\alpha^2-1\right)A+1,\quad\langle\psi_{\phi}\vert \psi_{\phi}\rangle=A,\quad\langle\psi_{\phi} \vert \psi_{\theta}\rangle=\alpha D,
\label{form17}
\end{eqnarray}
where
\begin{eqnarray}
&&A=\vert a_i\vert^2+\vert d_i\vert^2,\quad B=\vert b_i-c_i\vert^2,\quad D=\vert a_i\vert^2-\vert d_i\vert^2.
\label{form20}
\end{eqnarray}
Substituting these results into (\ref{form11}) and then into (\ref{form10}), we obtain
\begin{eqnarray}
&&ds^2=\gamma^2\left[\left(\alpha^2-1\right)A+1 -\left(\left(\alpha-1\right)A+1-B\right)^2\right](d\theta)^2\nonumber\\
&&+\gamma^2\left[A-D^2\right](d\phi)^2+2\gamma^2D\left[\alpha -\left(\left(\alpha-1\right)A+1-B\right)\right]d\theta d\phi.
\label{form18}
\end{eqnarray}
It is easy to see that components of the metric tensor do not depend on the parameters $\theta$ and $\phi$. This means that
the expression (\ref{form18}) defines the metric of a flat manifold. It is worth noting that this result is a
consequence of the commutativity between interaction and the magnetic-field parts of the Hamiltonian (\ref{form1}). Indeed,
let us consider the following unitary operator
\begin{eqnarray}
e^{-i(\theta H_1+\phi H_2)},
\label{add1}
\end{eqnarray}
where $H_1$ and $H_2$ are Hermitian operators which commute between themselves and do not depend on the mentioned parameters
$\theta$ and $\phi$. The action of this operator on the initial state $\vert\psi_i\rangle$ is as follows
\begin{eqnarray}
\vert\psi(\theta,\phi)\rangle= e^{-i\theta H_1}e^{-i\phi H_2}\vert\psi_i\rangle.
\label{add2}
\end{eqnarray}
Now, using the relations  (\ref{form10}) and (\ref{form11}) for this state we obtain the metric of manifold in the form
\begin{eqnarray}
&&ds^2=\gamma^2\left(\langle\psi_i\vert H_1^2\vert\psi_i\rangle-\langle\psi_i\vert H_1\vert\psi_i\rangle^2\right)(d\theta)^2+\gamma^2\left(\langle\psi_i\vert H_2^2\vert\psi_i\rangle-\langle\psi_i\vert H_2\vert\psi_i\rangle^2\right)(d\phi)^2\nonumber\\
&&+2\gamma^2 \left(\langle\psi_i\vert H_1H_2\vert\psi_i\rangle-\langle\psi_i\vert H_1\vert\psi_i\rangle\langle\psi_i\vert H_2\vert\psi_i\rangle\right)d\theta d\phi .
\label{add3}
\end{eqnarray}
Since the operators $H_1$ and $H_2$ do not depend on parameters $\theta$ and $\phi$, the metric (\ref{add3}) describes a flat manifold.

Also, it should be noted that the commutativity in the Hamiltonian (\ref{form1}) allows to calculate
the evolution in the case of a time-dependent magnetic field $h_z(t)$ easily. Then in the above results the parameter $\phi$
takes the form $\phi=2\int h_z(t)dt$. These results are important in applications of the model because the time-dependent magnetic
field additionally allows to control the evolution path on the manifold.

Let us analyze the geometry of the manifold defined by the expression (\ref{form18}) for the six cases considered in the previous section:
\begin{enumerate}
\item In the first case $A=D=0$ and
\begin{eqnarray}
ds^2=\gamma^2B(2-B)(d\theta)^2.
\label{form18_1}
\end{eqnarray}
Taking into account that $\theta\in[0,\pi]$ we conclude that this metric defines the circle with the radius $\frac{\gamma}{2}\sqrt{B(2-B)}$.
\item In this case $A=1$, $B=0$ and the metric takes a form
\begin{eqnarray}
ds^2=\gamma^2\left[1-D^2\right](d\phi)^2.
\label{form18_2}
\end{eqnarray}
This is the metric of the circle with radius $\frac{\gamma}{2}\sqrt{1-D^2}$, where $\phi\in[0,\pi]$.
\item In this case we obtain the metric of the circle
\begin{eqnarray}
ds^2=\gamma^2\left[A-D^2\right](d\phi)^2
\label{form18_3}
\end{eqnarray}
with radius $\gamma\sqrt{A-D^2}$ and $\phi\in[0,2\pi]$.
\item Analyzing periodic conditions (\ref{form5_9}) we conclude that in this case the manifold is a torus with $\phi\in[0,2\pi]$ and
$\theta\in[0,\frac{\pi}{\alpha\mp 1}]$. Also this manifold is twisted through an angle $\pi$ for the parameter $\phi$.
\item Here we also have a torus if $p$ and $q$ are both odd. If $p$ is even and $q$ is odd and vice versa we have a twisted torus
through an angle $\pi$ for a parameter $\phi$. For these two cases $\phi\in[0,2\pi]$ and $\theta\in[0,q\pi]$.
\item In this case we obtain the periodicity only for parameter $\phi\in[0,2\pi]$ because the parameter $\theta\in[-\infty,+\infty]$.
It means that manifold is closed along the parameter $\phi$ and
is open along $\theta$. We conclude that in this case it is an infinitely long cylinder.
\end{enumerate}

Also it is worth noting
that for $\alpha=1$, corresponding to the case of the isotropic Heisenberg interaction between spins, we obtain the metric
as in our previous paper \cite{torus}
\begin{eqnarray}
ds^2=\gamma^2\left[B\left(2-B\right)(d\theta)^2+\left(A-D^2\right)(d\phi)^2+2BDd\theta d\phi\right].
\label{form19}
\end{eqnarray}

\section{Entanglement of quantum states on a two-spin manifold \label{sec4}}

In the present section we consider entanglement of the states which can be achieved during the evolution of the system represented
by the Hamiltonian (\ref{form1}). Also we find conditions for the preparation of the maximally entangled quantum states on this system.
So, first of all we calculate entanglement of the states which belong to the manifold defined by the metric (\ref{form18}).
The degree of entanglement of two spin system can be determined by the concurrence \cite{ent1, ent2}
\begin{eqnarray}
C=2\vert ad-bc\vert,
\label{form32}
\end{eqnarray}
where parameters $a$, $b$, $c$ and $d$ are defined by expression (\ref{form4_1}).
Using this definition let us calculate the concurrence for the states (\ref{form5})
\begin{eqnarray}
C=2\vert a_id_ie^{-i2\alpha\theta}+\frac{i}{2}\left({b_i}^2+{c_i}^2\right)\sin 2\theta -b_ic_i\cos 2\theta\vert.
\label{form33}
\end{eqnarray}
It is easy to see that for a particular value of $\theta$ we can select the curves on manifold with a constant entanglement.
From the equation (\ref{form18}) we obtain that these curves are circles with radii depending on the parameters of the initial states as follows
\begin{eqnarray}
R=\gamma\beta\sqrt{A-D^2},
\label{form32_1}
\end{eqnarray}
where $\beta$ is a parameter that is defined by the periodicity of $\phi$. This parameter is equal to $1/2$ for case 2 of
the previous section and is equal to $1$ in all other cases except the first one. This is because in the first case we have the manifold
which does not depend on $\phi$. On this manifold, which is a circle, each point has another entanglement.

Let us consider entanglement of a two-spin system when the initial state is unentangled and has the form
\begin{eqnarray}
\vert\psi_i\rangle=\vert + -\rangle,
\label{form31_1}
\end{eqnarray}
where $\vert +\rangle = \cos\frac{\chi_i}{2}\vert\uparrow\rangle +\sin\frac{\chi_i}{2}e^{i\gamma_i}\vert\downarrow\rangle$,
$\vert -\rangle = -\sin\frac{\chi_i}{2}\vert\uparrow\rangle +\cos\frac{\chi_i}{2}e^{i\gamma_i}\vert\downarrow\rangle$. Here
parameters $\chi_i$ and $\gamma_i$ belong to the intervals $\chi_i\in\left[0,\pi\right]$ and $\gamma_i\in\left[0,2\pi\right]$,
respectively. It is worth noting that the state (\ref{form31_1}) can be easily prepared because it is the eigenstate of the system of two spins placed
in the strong magnetic field ($B\gg J$) which is directed along the unit vector ${\bf n}=\left[\sin\chi_i\cos\gamma_i,\sin\chi_i\sin\gamma_i,\cos\chi_i\right]$.
Substituting initial parameters from state (\ref{form31_1}) into the equation (\ref{form18}) we obtain that in this case the evolution happens
on the manifold with metric
\begin{eqnarray}
ds^2=\gamma^2\left[\left(\alpha^2-1\right)A+1 -\left(\alpha-1\right)^2A^2\right](d\theta)^2+\gamma^2A(d\phi)^2,
\label{form31_2}
\end{eqnarray}
where $A=\frac{1}{2}\sin^2\chi_i$. Now using the equation (\ref{form33}) with the state (\ref{form31_1}) we obtain that the concurrence
of the states belonging to this manifold takes the form
\begin{eqnarray}
C=\frac{1}{2}\left[\sin^4\chi_i\left(\cos 2\alpha\theta-\cos 2\theta\right)^2+\left((1+\cos^2\chi_i)\sin 2\theta+\sin^2\chi_i\sin2\alpha\theta\right)^2\right]^{1/2}.
\label{form33_0}
\end{eqnarray}
As we can see, the influence of $\alpha$ on the entanglement depends on the initial parameter $\chi_i$. The closer the
parameter $\chi_i$ is to the number $\pi/2$, the greater number $\alpha$ influences on the value of entanglement.
If $\chi_i=0,\pi$ then the concurrence does not depend on $\alpha$ and has the form
\begin{eqnarray}
C=\vert\sin 2\theta\vert.
\label{form33_1}
\end{eqnarray}
This is because the initial state has the form $\vert\uparrow\downarrow\rangle$ or $\vert\downarrow\uparrow\rangle$, respectively.
These states are the eigenstates of the $H_{zz}$ term of Hamiltonian (\ref{form1}), which contains the parameter $\alpha$.
We can see that for $\theta=\pi/4$ and $3\pi/4$ we obtain the states with maximum entanglement. For instance,
from the initial states $\vert\uparrow\downarrow\rangle$ modulo a global phase we obtain the following states
$\frac{1}{\sqrt{2}}\left(\vert\uparrow\downarrow\rangle-i\vert\downarrow\uparrow\rangle\right)$
for $\theta=\pi/4$ and
$\frac{1}{\sqrt{2}}\left(\vert\uparrow\downarrow\rangle+i\vert\downarrow\uparrow\rangle\right)$ for $\theta=3\pi/4$.

\begin{figure}[!!h]
\centerline{\includegraphics[scale=0.5, angle=0.0, clip]{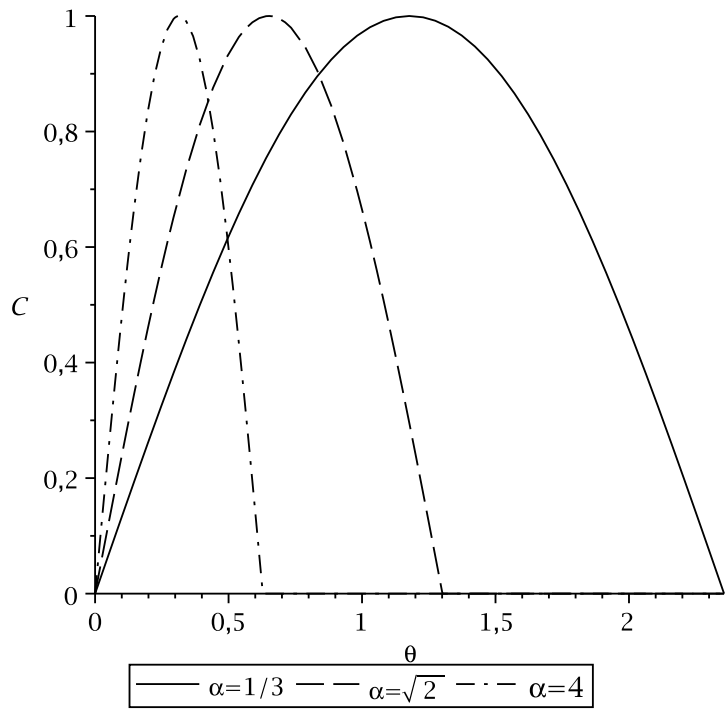}}
\caption{{\it The dependence of the concurrence on the angle $\theta$ (\ref{form33_2}) for three different $\alpha$. As we can see,
the greater $\alpha$, the faster the system reaches the maximally entangled state.}}
\label{ent}
\end{figure}

Otherwise, in the case of $\chi_i=\pi/2$ we obtain the maximal dependence of the concurrence on the parameter $\alpha$
\begin{eqnarray}
C=\vert \sin (\alpha+1)\theta\vert.
\label{form33_2}
\end{eqnarray}
As we can see, when $\alpha=-1$ then $C=0$. This is because the initial state
\begin{eqnarray}
\vert\psi_i\rangle =-\frac{1}{2}\vert\uparrow\uparrow\rangle+\frac{1}{2}e^{i\gamma_i}\vert\uparrow\downarrow\rangle-\frac{1}{2}e^{i\gamma_i}\vert\downarrow\uparrow\rangle+\frac{1}{2}e^{i2\gamma_i}\vert\downarrow\downarrow\rangle\nonumber
\end{eqnarray}
is an eigenstate of the Hamiltonian $H_{xx}+H_{zz}$ (see (\ref{form1_1}), (\ref{form1_2})) with the eigenvalue $-J$. In the case of $\alpha\neq-1$
the evolution happens on the manifold which is a torus with the properties described in the case 4 of the section \ref{sec3}. Taking into account that
$\theta\in\left[0,\pi/\alpha+1\right]$ it is easy to see that for
\begin{eqnarray}
\theta=\frac{\pi}{2(\alpha+1)}
\label{form33_3}
\end{eqnarray}
we have the maximal amount of the entanglement ($C=1$). Also taking into account that parameter $\theta$ and the period of evolution
are related by (\ref{form5_1}) we conclude that the greater the number $\alpha$, the faster the system
reaches the maximally entangled state. The dependence of the entanglement on the angle $\theta$ for three different $\alpha$
is shown in Fig. \ref{ent}. It is worth noting that the same results we obtain for the initial state $\vert-+\rangle$.

Finally, let us investigate entanglement when the initial state has the form $\vert++\rangle$ or $\vert--\rangle$.
Using eqution (\ref{form33}) we obtain that the concurrence for these cases has the form
\begin{eqnarray}
C=\vert\sin^2\chi_i\vert \vert \sin (\alpha-1)\theta\vert.
\label{form33_4}
\end{eqnarray}
Analysing the relation (\ref{form33_4}) we can conclude that the maximally entangled state ($C=1$) is achieved only for $\chi_i=\pi/2$ and when $\alpha\neq1$. Here the
evolution also happens on the manifold which is a torus. Its geometrical properties are described in the case 4 of section \ref{sec3}.

\section{The quantum evolution of two spins with DM interaction \label{sec5}}

In this section we consider a two-spin system described by DM and $ZZ$ interactions placed in the
magnetic field
\begin{eqnarray}
H'=H_{DM}+H_{zz}+H_{mf},
\label{form6}
\end{eqnarray}
where
\begin{eqnarray}
H_{DM}=J\left(\sigma_x^1\sigma_y^2-\sigma_y^1\sigma_x^2\right).\label{form1_3}
\end{eqnarray}
This Hamiltonian has four eigenvalues: $\alpha J+2h_z$, $\alpha J-2h_z$,
$-\alpha J+2J$, and $-\alpha J-2J$ with the corresponding eigenvectors
\begin{eqnarray}
&&\vert\uparrow\uparrow\rangle,\label{form6_1}\\
&&\vert\downarrow\downarrow\rangle,\label{form6_2}\\
&&\frac{1}{\sqrt{2}}\left(\vert\uparrow\downarrow\rangle - i\vert\downarrow\uparrow\rangle\right),\label{form6_3}\\
&&\frac{1}{\sqrt{2}}\left(\vert\uparrow\downarrow\rangle + i\vert\downarrow\uparrow\rangle\right).
\label{form6_4}
\end{eqnarray}
The Hamiltonian (\ref{form6}) can be obtained from the Hamiltonian (\ref{form1}) making the following
unitary transformation
\begin{eqnarray}
H'=e^{i\frac{\pi}{4}\sigma_z^1}He^{-i\frac{\pi}{4}\sigma_z^1}.
\label{form6_5}
\end{eqnarray}
This transformation reflect the term $H_{XX}$ in the Hamiltonian (\ref{form1}) into the term $H_{DM}$ and does not change other terms of this Hamiltonian.
Using this fact let us represent the operator of evolution with Hamiltonian (\ref{form6}) in the form
\begin{eqnarray}
U'(t)=e^{i\frac{\pi}{4}\sigma_z^1}e^{-iH_{xx}t}e^{-i\frac{\pi}{4}\sigma_z^1}e^{-iH_{zz}t}e^{-ih_z\sigma_z^1t}e^{-ih_z\sigma_z^2t},
\label{form7}
\end{eqnarray}
where operators $e^{-iH_{xx}t}$ and $e^{-iH_{zz}t}$ are represented by equations (\ref{form3_1}) and (\ref{form3_2}), respectively.
In the basis labelled by
$\vert\uparrow\uparrow\rangle$, $\vert\uparrow\downarrow\rangle$,
$\vert\downarrow\uparrow\rangle$ and
$\vert\downarrow\downarrow\rangle$, operator (\ref{form7}) can be represented as follows
\begin{eqnarray}
U'(t)=\left( \begin{array}{ccccc}
e^{-i\left(2h_z+\alpha J\right)t} & 0 & 0 & 0\\
0 & \cos\left(2J t\right)e^{i\alpha Jt} & \sin\left(2J t\right)e^{i\alpha Jt} & 0 \\
0 & -\sin\left(2J t\right)e^{i\alpha Jt} & \cos\left(2J t\right)e^{i\alpha Jt} & 0 \\
0 & 0 & 0 & e^{i\left(2h_z-\alpha J\right)t}
\end{array}\right).
\label{form8}
\end{eqnarray}
The state, which is the result of the evolution of the two-spin system having started from the initial state (\ref{form4_1}) takes the form
\begin{eqnarray}
&&\vert\psi'(\theta,\phi)\rangle=e^{i\frac{\alpha\theta}{2}}\left[a_ie^{-i\left(\phi+\alpha\theta\right)}\vert\uparrow\uparrow\rangle + \left(b_i\cos\theta+c_i\sin\theta\right)\vert\uparrow\downarrow\rangle \right.\nonumber\\
&&\left. + \left(-b_i\sin\theta+c_i\cos\theta\right)\vert\downarrow\uparrow\rangle + d_ie^{i\left(\phi-\alpha\theta\right)}\vert\downarrow\downarrow\rangle\right].
\label{form9}
\end{eqnarray}
Here we introduce the same notations (\ref{form5_1}) as we have used for the state (\ref{form5}). This state is also defined by two real parameters $\theta$ and
$\phi$. The periodic conditions, obtained in the section \ref{sec2}, are satisfied by the state (\ref{form9}) if we replace $b_i$
on $-ib_i$ or $c_i$ on $ic_i$.

The Fubini-Study metric of the manifold defined by the state (\ref{form9}) can be calculated using the fact that
the relationship between the Hamiltonians (\ref{form1}) and (\ref{form6}) is defined by expression (\ref{form6_5}).
The metric of this manifold is determined by the expression (\ref{form18}) with $B=\vert b_i-ic_i\vert^2$.
The geometry properties of this manifold can be determined similarly as in the section \ref{sec3}, using corresponding periodic conditions.

\begin{figure}[!!h]
\centerline{\includegraphics[scale=0.5, angle=0.0, clip]{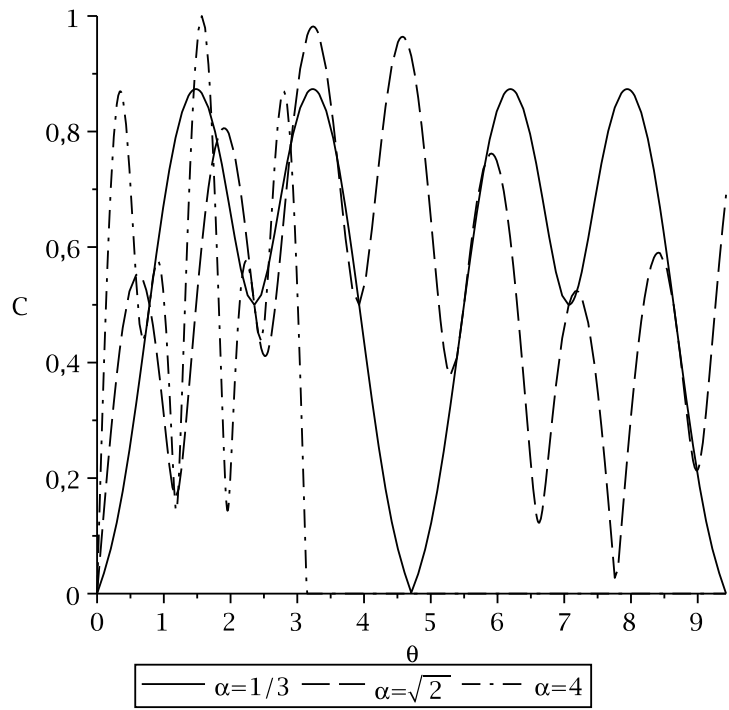}}
\caption{{\it Dependence of the concurrence on the angle $\theta$ (\ref{form42}) for three different $\alpha$ in the case of
two spins described by the Hamiltonian (\ref{form6})}}.
\label{DMent}
\end{figure}

Let us consider the entanglement of this system when the initial state is determined by the equation (\ref{form31_1}). In this case we obtain
the following expression for the concurrence
\begin{eqnarray}
C=\frac{1}{2}[\sin^4\chi_i\sin^22\alpha\theta+(\sin^2\chi_i(\cos2\theta-\cos 2\alpha\theta)+2\cos\chi_i\sin 2\theta)^2]^{1/2}.
\label{form41}
\end{eqnarray}
As we can see influence of parameter $\alpha$ on the behavior of entanglement increases when the value of parameter $\chi_i$
tends to $\pi/2$. If $\chi_i=0,\pi$ then concurrence is defined by the equation (\ref{form33_1}). For example, for $\chi_i=\pi$
we obtain modulo a global phase the maximally entangled Bell states (\ref{form2_3}) for $\theta=\pi/4$ and (\ref{form2_4})
for $\theta=3\pi/4$. If $\chi_i=\pi/2$ the concurrence takes the form
\begin{eqnarray}
C=\frac{1}{2}[1+\cos2\theta(\cos 2\theta-2\cos 2\alpha\theta)]^{1/2}.
\label{form42}
\end{eqnarray}
From the analysis of this expression it is clear that the time required for achieving the maximally entangled states depends
on $\alpha$. Similarly as in the case of anisotropic Heisenberg interaction we can conclude that the greater $\alpha$,
the faster the system reaches maximally entangled state (Fig. \ref{DMent}).

\section{Conclusion \label{sec6}}

We studied the quantum evolution of a two-spin system represented by the anisotropic Heisenberg Hamiltonian
in the magnetic field directed along the $z$-axis. It is defined by two real parameters,
namely, the period of the time of evolution and the value of the magnetic field. This means that the system evolves on a manifold
determined by these parameters. We calculated the Fubini-Study metric of the manifold and showed that it is flat.
The geometry and the size of this manifold depends on the ratio between interaction couplings of the Hamiltonian and on the parameters
of the initial state. The entanglement of the states belonging to this manifold was investigated. It was found that the curves of a
constant entanglement are circles with radii depending on the initial states. In the case of the unentangled
initial state (\ref{form31_1}) the dependence of the entanglement of the system on the ratio between interaction couplings was obtained.
We showed that the greater the ratio, the faster the system reaches maximally entangled state.
If the initial state is directed along the $z$-axis this dependence disappears.

Finally, we considered the evolution of a two-spin system described by DM interaction with $ZZ$ one.
Due to the fact that this Hamiltonian and anisotropic Heisenberg Hamiltonian (\ref{form1}) are linked by the unitary
transformation (\ref{form6_5}) we calculated the Fubini-Study metric of the manifold which defines such evolution. Similarly
as in the case of anisotropic Heisenberg interaction we obtained that this metric describes a flat manifold with geometry depending
on the ratio between interaction coupling and on the initial state.

\section{Acknowledgement}

First of all, the author thanks Prof. Volodymyr Tkachuk for his great support during the investigation of the problem.
Also the author thanks Drs. Taras Krokhmalskii, Andrij Rovenchak and Mykola Stetsko for useful comments. This work was partly supported
by Project FF-30F (No. 0116U001539) from the Ministry of Education and Science of Ukraine and by the grant from
the State Fund For Fundamental Research of Ukraine, Competition F-64/41-2015 (No. 0115U004838),
Project "Classical and quantum systems beyond standard approaches".

\end{document}